\documentclass{article}
\usepackage{graphicx} 
\usepackage{courier}
\usepackage{tcolorbox}
\usepackage{titlesec}
\usepackage{changepage}
\usepackage[a4paper, margin=1in]{geometry}
\usepackage{booktabs}
\usepackage{graphicx}
\usepackage{listings}
\usepackage{xcolor}
\definecolor{lightforestgreen}{rgb}{0.13, 0.55, 0.13} 
\definecolor{lightergray}{rgb}{0.9, 0.9, 0.9} 

\lstdefinestyle{mystyle}{
    backgroundcolor=\color{lightergray},   
    basicstyle=\ttfamily\footnotesize\color{lightforestgreen},  
    frame=none,                      
    breaklines=true,
}

\title{The Use of Large Language Models (LLM) for Cyber Threat Intelligence (CTI) in Cybercrime Forums}

\author{Vanessa Clairoux-Trépanier\textsuperscript{1}\footnote{These two authors contributed equally to the study},
Isa-May Beauchamp\textsuperscript{1}*
,
Estelle Ruellan\textsuperscript{2},\\
Masarah Paquet-Clouston\textsuperscript{1,3},
Serge-Olivier Paquette\textsuperscript{2},
Eric Clay\textsuperscript{2}}

\date{White Paper, October 2024}

\begin{document}

\maketitle
\begin{adjustwidth}{0.5in}{0.5in}
\noindent
\hspace{0.05in}\textbf{1} School of Criminology, Université de Montréal, Montréal, Québec, Canada
\\
\vspace{0.1cm}
\noindent
\textbf{2} Flare Systems, Canada
\\
\vspace{0.1cm}
\noindent
\textbf{3} Complexity Science Hub, Vienna, Austria
\\
\bigskip
\end{adjustwidth}

\begin{center}
\textbf{\Large Abstract}
\end{center}
\begin{adjustwidth}{0.5in}{0.5in}
Large language models (LLMs) can be used to analyze cyber threat intelligence (CTI) data from cybercrime forums, which contain extensive information and key discussions about emerging cyber threats. However, to date, the level of accuracy and efficiency of LLMs for such critical tasks has yet to be thoroughly evaluated. Hence, this study assesses the performance of an LLM system built on the OpenAI GPT-3.5-turbo model~\cite{openai_gpt3_5_turbo} to extract CTI information. To do so, a random sample of more than 700 daily conversations from three cybercrime forums—XSS, Exploit.in, and RAMP—was extracted, and the LLM system was instructed to summarize the conversations and predict 10 key CTI variables, such as whether a large organization and/or a critical infrastructure is being targeted, with only simple human-language instructions. Then, two coders reviewed each conversation and evaluated whether the information extracted by the LLM was accurate. The LLM system performed well, with an average accuracy score of 96.23\%, an average precision of 90\% and an average recall of 88.2\%. Various ways to enhance the model were uncovered, such as the need to help the LLM distinguish between stories and past events, as well as being careful with verb tenses in prompts. Nevertheless, the results of this study highlight the relevance of using LLMs for cyber threat intelligence.
\end{adjustwidth}

\section{Introduction}
With the rise of large language models (LLMs), which are text-generating technologies trained on vast amounts of data~\cite{ferrag_generative_2024,hassanin_moustafa_2024, xu2024large}, the scope and aim of artificial intelligence (AI) have changed. Now, AI can be used for a myriad of applications, such as writing stories on the fly or summarizing complex scientific content~\cite{sheikh_artificial_2023}. One key application where AI can be useful is cyber threat intelligence (CTI)~\cite{ferrag_generative_2024,hassanin_moustafa_2024, xu2024large}. Indeed, there exists a vast array of cybercrime conversations in forums on both the clear and dark web~\cite{adewopo_exploring_2020, liggett_dark_2020}. These conversations often leak key information that can be used by companies and governments to detect and sometimes even prevent cyber attacks~\cite{adewopo_exploring_2020, liggett_dark_2020}. The question is whether and how LLMs can accurately be used for cyber threat intelligence on such forums. Indeed, can we trust such technology for CTI? Can it replace first-level threat analysts, that is, analysts who read and extract relevant information from cybercrime forums? And at what cost?

This study assesses the extent to which an LLM system is accurate when extracting and summarizing information from cybercrime forums. Using a random sample of more than 700 daily conversations from three cybercrime forums—XSS, Exploit[.]in, and RAMP—we instructed an LLM to summarize the conversations and extract specific CTI information from them. CTI information included whether a sale was conducted, whether a large organization or a critical infrastructure was discussed, whether initial access to an organization was mentioned, whether a vulnerability that is remotely exploitable or actively exploited was mentioned, and whether users discussed geopolitical conflicts. Any technology mentioned in a conversation was also coded, as well as any industry, when available. Such information is key for CTI: it allows analysts to narrow down cybercrime conversations that focus, for example, on a specific technology in an industry or on specific vulnerabilities targeting large organizations. 

It is worth mentioning as well that traditional machine learning classifiers meant to do that with extensive and expensive training data that were manually labelled and curated. The aim of this study is to assess the capacity of LLMs to perform these task without having to prepare such a training corpus, in a so-called \textit{zero-shot} setting.

To assess the performance of the LLM, two coders conducted a thorough review of each conversation and evaluated whether the information extracted by the LLM was accurate. This process 1) determined the level of performance of the technology according to accuracy, precision and recall, and 2) highlighted flaws that, once acknowledged, could be fixed.

In the end, the LLM system performed well, with an average accuracy score of 96.23\%, an average precision of 90\% and an average recall of 88.2\%. The assessment of the LLM also highlighted areas of improvement that could be useful for any researcher wishing to use LLMs to summarize forum conversations, including verb tenses in prompts and the impact of using large or vague concepts. The results validate that large language models can be effectively used for cyber threat intelligence.

\section{Methods and Data}

The following sections describe the LLM system developed for this study, the CTI variables it coded, and the manual process used to verify the results.

\subsection{LLM System for CTI}

For this study, the LLM system, powered by the gpt-3.5-turbo-16k-0613 model~\cite{openai_gpt3_5_turbo_0613, openai_gpt3_5_turbo}, is designed to extract and summarize relevant information from cybercrime forums. The system operates through a multi-step process that involves selecting high-quality sources, summarizing conversations, and coding key variables. This section outlines the detailed methodology employed in generating unit summaries, including pseudo-code and an example of a modular prompt.

\subsubsection{Data Collection and Preprocessing}

The first step involves selecting and collecting data from high-quality cybercrime forums. The system processes, every day, new daily conversations from new and existing threads on these cybercrime forums. These daily conversations (also called events) are then analyzed.

\subsubsection{Contextual Information Extraction}
For each new daily extractions, the system determines the context based on whether the discussion thread is new or has been previously processed. If the thread is new, the title serves as the context. For existing threads, a summary of the prior conversations is used to provide context.

\subsubsection{Prompt Design and Variable Extraction}
The core of the system's functionality lies in the use of carefully designed prompts that guide the LLM to extract specific information. Prompts are formulated to capture the intent and requirements for data extraction, such as identifying transactions, potential targets, or vulnerabilities mentioned in the conversation. The prompts are designed from the perspective of a cyber threat intelligence analyst, focusing on the key elements that are critical for threat analysis and reporting. The system's output is structured as unit summaries, which include a text summary of the conversation and the extracted variables in a standardized format. This format facilitates easy analysis and integration into further processing pipelines. We explicitly instruct the LLM to output the information in the specified format.

\subsubsection{Example Modular Prompt with Analyst Persona}

\begin{adjustwidth}{0.5in}{0.5in}
\begin{lstlisting}[style=mystyle]

"As a cyber threat intelligence analyst, your task is to review the conversation and identify key indicators. Please extract the following information: 1. An actor is selling something? 2. The conversation involves the sale of initial access to a corporate or organization network? 3. Targeted or abused products or technology names?". Output your answer in the following format, as an example.

{
 "summary": "The conversation focused on the sale of a new exploit targeting XYZ software. An actor claimed to have discovered a vulnerability and is offering it for sale. There was also a discussion about potential targets using ABC technology.",
 "variables": {
     "is_sale": true,
     "is_initial_access": true,
     "targeted_technologies": "XYZ software, ABC technology"
 }
}
\end{lstlisting}
\end{adjustwidth}

The following pseudo-code outlines the steps taken to generate unit summaries:

\begin{adjustwidth}{0.5in}{0.5in}
\begin{lstlisting}[style=mystyle]
    FOR each message in daily_batch in the thread DO
    context = daily_batch
    IF thread is new THEN
    context = extract_thread_title(message) + context
    ELSE context += retrieve_thread_summary(message) + context
    END IF 
    response = LLM.generate_response(context, PROMPT)
    unit_summary = {
    "summary": extract_summary(response),
    "variables": extract_variables(response) }
    STORE unit_summary 
    END FOR
\end{lstlisting}
\end{adjustwidth}

In this pseudo-code, PROMPT represents the modular prompt guiding the LLM system, and \texttt{\textcolor{lightforestgreen}{generate\\\_response}}, \texttt{\textcolor{lightforestgreen}{extract\_summary}}, and \texttt{\textcolor{lightforestgreen}{extract\_variables}} are functions that interact with the LLM model to produce the required outputs.

\subsection{Key CTI Variables and Prompts}

From each event, ten variables were extracted using specific but simple prompts with the LLM. These variables were chosen because they provide key and actionable information for CTI, such as which technology or industry is targeted. Having readily accessible and easily identifiable information for CTI ensures that analysts can narrow down their focus to conversations relevant for their organization, or get alerted when matching certain criteria. Anyone wishing to reproduce the analysis could develop their own key variables. Table~\ref{table:cybercrime_prompts} presents the variables that were extracted by the LLM for this study and their related prompts.

    \begin{table}[h!]
        \centering
        \begin{tabular}{|l|p{9cm}|}
            \hline
            \textbf{Variable Name} & \textbf{Prompt} \\
            \hline
            summary (char) & Generate concise extraction summaries with ample information for effective similarity searches, including conversation outcomes and all technical details that could be used by an analyst to investigate the threat. \\
            \hline
            is\_sale (1/0) & An actor is selling something. \\
            \hline
            is\_initial\_access (1/0) & Involves the sale of initial access to corporate or organization network. \\
            \hline
            is\_targeting\_large\_organization (1/0) & A large organization is being targeted by the actors. \\
            \hline
            is\_targeting\_critical\_infrastructure (1/0) & A critical infrastructure provider is targeted by the threat. \\
            \hline
            is\_remotely\_exploitable (1/0) & A mentioned vulnerability is remotely exploitable. \\
            \hline
            is\_actively\_exploitable (1/0) & A mentioned vulnerability is being actively exploited. \\
            \hline
            is\_geopolitics (1/0) & The discussion involves geopolitical issues. \\
            \hline
            targeted\_technologies (list) & Targeted or abused products or technology names. \\
            \hline
            industries (list) & Names of the industries relevant to the content, including those targeted by threat actors. Choose from the following options, each accompanied by a specific definition: \\ 
            & \textit{Finance}: Involving banking, cryptocurrencies, investment, insurance, real estate, stock market, money mule, embezzlement, money laundering, insider trading, shell companies, and other financial services. \\
            & \textit{Technology and Software}: Involving software development, cryptocurrencies, blockchain, IT services, hardware manufacturing, electronics, and related fields. \\
            & \textit{Critical Infrastructure}: Covering essential systems and facilities such as energy, oil and gas sector, transportation, water supply, telecommunications, internet providers, military, governments, harbor, airport. \\
            & \textit{Healthcare}: Involving medical services, hospitals, clinics, pharmaceuticals, biotechnology, emergency centers, and healthcare IT. \\
            & \textit{Other}: Any industry not explicitly mentioned above. \\
            & \textit{All}: Indicates that the content may be relevant to all industries. \\
            \hline
        \end{tabular}
        \caption{Key information from cybercrime conversations and associated prompts}
        \label{table:cybercrime_prompts}
    \end{table}

\subsection{Coding Process}
To assess the accuracy of the LLM system, a random sample of 500 daily conversations from three cybercrime forums—XSS, Exploit.in, and RAMPs— were extracted using the Flare interface. Flare is an information technology (IT) security company that maintains a cyber threat intelligence platform by monitoring various online spaces\footnote{https://flare.io/}. 

The LLM model discussed above was then applied on these conversations. However, some variables had unbalanced results (i.e., had a small number of TRUE instances). To overcome this limit, we randomly selected additional conversations to make sure that each binary variable had at least 50 conversations scoring TRUE and 50 conversations scoring FALSE. 

Then, two analysts went over each of the daily conversations from cybercrime forums and assessed the accuracy of the summary coded by the LLM system. They also individually coded the ten variables according to their interpretation, then compared them to the coding performed by the LLM. Specifically, they used a binary coding scheme: 1 indicated agreement with the LLM coding and 0 indicated disagreement.

Once the individual coding was completed, the two analysts performed an inter-coder agreement to obtain a common decision for each unit summary. Such inter-coder agreement allowed them to pinpoint discrepancies, but also find areas of improvement to fine-tune the LLM system. In the end, the inter-coder agreement was high, with an average of 95.05\%. A merged database was created resulting from the agreement. This merged database was used to determine the performance, in terms of accuracy, precision and recall, of the LLM system in coding each of the CTI variables from the daily conversations on cybercrime forums. The results are presented below.

\subsection{Accuracy, Precision and Recall}

Accuracy measures the proportion of correct predictions (both true positives (TP) and true negatives (TN)) out of the total prediction~\cite{hastie2009elements}. It is used as a primary metric for predictive systems. However, in case of imbalanced datasets, where the majority class vastly outweighs the minority, accuracy can be misleading. A system might achieve high accuracy by focusing on the majority class while missing important minority class instances, which are often the most critical in cybersecurity. This is where precision and recall offer more insight into the system’s performance~\cite{hastie2009elements}.

To give an example, take the variable \texttt{is\_remotely\_exploitable} sampled in a way that eight messages out of 500 are set to TRUE. Then, a system predicting FALSE 100\% of the time would yield an accuracy of 492/500 = 98.4\%, but the system is effectively useless as it can’t be used to detect remotely exploitable vulnerabilities as part of messages. Several solutions exist in order to better assess the true ability of the model to detect the needed signal.

Precision and Recall are two complementary metrics that offer a more nuanced understanding of the underlying phenomena, especially when working with imbalanced datasets. Precision is defined as the proportion of correctly identified true instances out of all instances the system predicted as positive. In contrast, recall refers to the system’s ability to capture all actual true cases from the dataset.

Precision is often used when predictions are costly or risky, and when the cost of ignoring false negatives (FN) predictions is lesser than the cost of predicting the false elements as being true. Recall, on the other hand, is often used when the cost of missing a prediction is very high, for example, in cancer detection or in detecting active cyber threats~\cite{hastie2009elements}.

In cybersecurity, especially when dealing with real-life threat detection where the consequences of missing threats or mis-identifying signals can be extreme, these metrics are necessary to correctly understand the actual performance and reliability of the system. A system that prioritizes precision over recall may be suitable when false positives (FP) are highly undesirable, as could be the case when investigating threat actors in cybercrime forums. On the other hand, high recall is critical when it is more important not to miss any single threats, even at the cost of a potentially large number of false positives (FP), as in a critical infrastructure security operations environment.

In our context of CTI variable extraction, below is an interpretation of the performance of the LLM system tested. Note that since the original scope of the research was to identify geopolitical conversations in cybercrime forums, we first selected a balanced sample of the \texttt{is\_geopolitics} variable and then assessed the system’s extraction capability for the other variables on this sample, leading to the varying levels of class imbalance shown below.

\section{Assessing the Performance of the LLM System for CTI}

The LLM system performed well in coding variables that represented valuable CTI information from cybercrime forums. Indeed, on average, the LLM was accurate in 96.23\% of cases, with a minimum accuracy of 92.2\% for the variable \texttt{targeted\_technologies} and a maximum of 98.9\% for the variable \texttt{industries}. For binary variables, the average precision was 90\%, with a minimum of 75\% and a maximum of 96.1\%. The average recall was 88.2\% with a minimum of 67.6\% and a maximum of 99.2\%. The percentage performance metrics for each variable are presented in Table~\ref{table:binary_results} for the binary variables and in Table~\ref{table:char_results} for character variables.

\begin{table}[ht]
    \centering
    \resizebox{\textwidth}{!}{%
    \begin{tabular}{lrrrrrrrrr}
    \toprule
        Column &  N & Precision &  Recall  &  Accuracy  &  TP &  TN &  FP &  FN \\
    \midrule
        is\_geopolitics & 617 & 0.922  & 0.976  & 0.956 & 247 & 343 & 21 & 6 \\
         is\_sale & 702 & 0.909 & 0.992 & 0.962 & 250 & 425 & 25 & 2 \\
        is\_initial\_access & 617 & 0.961 & 0.952 & 0.985 & 99 & 509 & 4 & 5 \\
        is\_targeting\_large\_organizations & 702 & 0.957 & 0.815 & 0.949 & 132 & 534 & 6 & 30 \\
        is\_targeting\_critical\_infrastructure & 617 & 0.852 & 0.676 & 0.927 & 69 & 503 & 12 & 33 \\
        is\_remotely\_exploitable & 702 & 0.949 & 0.822 & 0.972 & 74 & 608 & 4 & 16 \\
        is\_actively\_exploited & 702 & 0.750  & 0.941 & 0.973 & 48 & 635 & 16 & 3 \\
    \bottomrule
    \end{tabular}%
    }
    \caption{Performance Results for Binary Variables}
    \label{table:binary_results}
\end{table}

\begin{table}[h!]
    \centering
    \begin{tabular}{|l|r|r|}
        \hline
        \textbf{Variable Name} & \textbf{N} & \textbf{Accuracy} \\
        \hline
        summary & 500 & 0.988\\
        \hline
        targeted\_technologies & 231 & 0.922 \\
        \hline
        industries & 365 & 0.989 \\
        \hline
    \end{tabular}
    \caption{Performance for Character Variables}
    \label{table:char_results}
\end{table}

For the \texttt{summary} variable, the coders ensured that the summary provided by the LLM system was accurate. The summary was accurate 98.8\% (N=500) of the time. However, coders occasionally noticed missing key CTI information from the summaries. This is likely due to the inherent nature of summarization, which involves omitting some details. Also, what is considered essential can vary between coders.

The variable \texttt{is\_sale} was coded well by the LLM system, achieving a precision of 90.9\% and a recall of 99.2\%. This means that the system accurately identified sales-related conversations with few false positives (precision), while nearly all true sales conversations were detected (high recall). Errors occurred when the LLM mistakenly flagged discussions about sales interest or general sales-related talk as actual sales. The variable \texttt{is\_initial\_access} performed similarly well, with a precision of 96.1\% and recall of 95.2\%, capturing almost all cases of initial access sales. The high precision and recall suggest that the LLM is reliable in detecting conversations around sales and access-related activities in cybercrime forums.

The variable \texttt{is\_targeting\_large\_organization} had a precision of 95.7\% but a lower recall of 81.5\%, indicating that while most flagged conversations were correctly identified as targeting large organizations, the LLM missed some relevant conversations (lower recall). Similarly, the variable  \texttt{is\_targeting\_critical\_infrastructure} had a precision of 85.2\% and a lower recall of 67.6\%. The errors in these two variables often stemmed from difficulties in defining what qualifies as a ``large organization" or a ``critical infrastructure." Despite these issues, analysts can still use these variables to get an overall sense of whether large organizations or critical infrastructures are being discussed in cybercrime forums.

The variable \texttt{is\_remotely\_exploitable} flagged conversations about remotely exploitable vulnerabilities, with a precision of 94.9\% and recall of 82.2\%. This suggests that most flagged vulnerabilities were relevant (high precision), but some vulnerabilities were missed (lower recall). On the other hand, the variable \texttt{is\_actively\_exploited} achieved a precision of 75\% and recall of 94.1\%, meaning that while some false positives were generated (lower precision), most conversations about actively exploited vulnerabilities were detected (high recall). Together, these two variables can help analysts identify vulnerabilities that organizations should prioritize for remediation.

The variable \texttt{is\_geopolitics} flagged conversations related to geopolitical issues and was coded with a precision of 92.2\% and recall of 97.6\%. This high recall means that the LLM captured almost all relevant conversations. However, the precision was slightly lower because the LLM tended to flag the variable as \texttt{TRUE} whenever a country was mentioned, even when the discussion wasn't strictly geopolitical. In this case, analysts can use this variable to focus on conversations that may impact specific countries or geopolitical conflicts.

For the variable \texttt{targeted\_technologies}, coders ensured that the technologies mentioned in messages matched those identified by the LLM. With an accuracy of 92.2\%, the LLM rarely missed relevant technologies but occasionally flagged some incorrectly. Given this high accuracy, analysts can confidently use this variable to monitor which technologies are being discussed in cybercrime forums.

Finally, the variable \texttt{industries} achieved an accuracy of 98.9\%, meaning that almost every industry mention was correctly identified without any false positives (FP) or false negatives (FN). This high score might be due to the better-defined nature of this variable, contributing to the LLM’s effectiveness in identifying industry mentions. Analysts and policymakers can use this variable to track discussions in cybercrime forums that target specific industries, facilitating better risk assessments.

In conclusion, the differences between precision and recall emphasize the interpretative nature of the variables. Precision reflects how accurately the LLM identifies relevant cases without false positives (FP), while recall shows its ability to capture all relevant instances. High precision and recall for variables like \texttt{is\_sale} and \texttt{is\_initial\_access} indicate reliable performance. However, variables like \texttt{is\_targeting\_critical\_infrastructure} have lower recall, meaning relevant conversations were sometimes missed, highlighting the need for improvement in capturing more nuanced discussions, particularly around complex definitions like ``critical infrastructure."

\section{Coders’ Insights to Fine-Tune the LLM System}
Given the results of the analysis, there is no doubt that LLM systems can be useful in extracting key CTI information from cybercrime forums. Indeed, the summaries generated from daily conversations focused on relevant information, even when a large number of messages were posted. Such relevant information was sometimes even missed by the analysts. 

Reviewing the same conversations allowed the coders to discuss and identify areas for improvement in coding CTI variables. These areas are useful for any researcher or analyst wishing to use LLMs to code variables based on text input. They are presented below.

\subsection{Difficulties in Detecting Stories and Past Events}

Across variables, the small number of observations that were miscoded often involved stories or past events mentioned by a user. Indeed, the LLM sometimes encountered difficulties when processing past events reported by users. For example, the LLM system coded the variable \texttt{is\_sale} as \texttt{TRUE} for a conversation flow that was summarized as follows:

\begin{quote}
``A police officer was caught selling fake certificates through a darknet market. The price per certificate was \$9,000, and during a bulk sale of \$80,000 worth of certificates, the officer attempted to deliver them personally, leading to his arrest." (382)
\end{quote}

However, in this message, the user was reporting a past event. Consequently, the LLM incorrectly categorized this example as a case of a user selling certificates rather than understanding it as a narrative of a past event involving a third party. Similarly, the LLM coded the variable \texttt{is\_sale} as \texttt{TRUE} for a conversation that was summarized as follows: 
\begin{quote}
    ``[An actor] reported that Russian national Evgeny Doroshenko has been charged in the USA for working as an initial access broker. Doroshenko is suspected of hacking at least one company in New Jersey and has been providing similar services since 2019. He set the starting price for access to the compromised company at \$3,000, with an auction increment of \$500 or an instant sale price of \$6,000. His preferred attack method is brute-forcing Remote Desktop Protocol services. Doroshenko's personal information, including phone numbers and home address in Astrakhan, was found linked to his Telegram account. Following the news of the charges, he attempted to contact moderators on the Exploit forum." (275)
\end{quote}

Again, in this conversation, a past event was reported, and therefore, the LLM should not have coded it as an active sale. When the LLM incorrectly coded such messages as \texttt{TRUE}, it over-represents the number of sales in the dataset. Hence, it is important to enhance the LLM's ability to distinguish historical narratives from current facts to ensure a more accurate and contextually appropriate analysis of the information provided by users.

\subsection{Verb Tenses in Prompts}
The coders also noticed that, given that the prompts were written in the present tense and the unit summaries were written in the past tense, the LLM sometimes had difficulties in coding accurately. For example, the variable \texttt{is\_sale} was not coded as \texttt{TRUE} for the conversation flow related to this summary:

\begin{quote}
``Two actors participated in a conversation. [the first actor] posted a message that was likely truncated and is not informative on its own. Actor [2]  indicated that a previous discussion or sale has been closed and the item in question has been sold. No further details are provided." (60)
\end{quote}

Most likely, the verb tense in the prompt, ``an actor is selling something," was the reason why the LLM did not code this conversation as involving a sale. However, upon reflection, it could be considered a sale because the item in question has been sold. In the end, if the goal is to code all sales, past or present, it might be necessary to revisit the verb tenses in the prompts given to the LLM.

\subsection{The Importance of Data Chunking}

During the coding process, the coders went through the \texttt{summary} variables and then the whole conversation on the cybercrime forum to validate that no information was missing and subsequently assess the accuracy of the CTI variables. When reading the complete conversation, the coders noticed that the way the data was chunked influenced the results. For example, given this summary:

\begin{quote}
``No actionable intelligence was extracted from the new message by [an actor] dated [xx-xx-xx] as it contained no specific information related to the concepts of interest for threat intelligence." (165)
\end{quote}

Almost all variables were coded as \texttt{FALSE}. However, when reading the complete conversation, one could quickly see that the discussion revolved around selling a database containing various Telegram chats and channels. The main user made various updates on the database's availability and scope, including traffic sources and promotional offers. However, the posts spanned many days. Hence, although the LLM’s result was accurate, the way the data was chunked led to the conclusion that the conversation did not revolve around a sale, although it did.

It is important to consider that how data is chunked and fed to the LLM may lead to an under-representation of some variables, as in the example above, or an over-representation if multiple ``pieces" of the same thread exposing one single sale, for example, were coded in several places. One might wonder whether taking the whole thread would be more efficient than daily conversations. Such a decision may depend on the length of the threads, as threads that span many days may lead to various discussion avenues. In the end, such observations have to be taken into account when coding the LLM system.

\subsection{About Coding General or Vague Concepts}
When coding CTI information, some variables relate to specific well-defined concepts, such as the variable \texttt{industries}. Indeed, the prompt for \texttt{industries} was well defined and included clear descriptions of industries, such as finance, technology, or critical infrastructure. On the other hand, other variables referred to general or vague concepts. For example, the coders noticed that some organizations that could be considered ``large", and therefore coded as TRUE for the variable \texttt{is\_targeting\_large\_organization}, were not coded as such by the LLM system. Since no definition was provided for what constitutes a ``large organization," this variable was subject to the LLM's interpretation. This means that only organizations recognized as such by the LLM were identified, or —as the coders noticed— only without-a-doubt large organizations were flagged (e.g., Apple, Netflix, Microsoft). Large organizations that were less known or required additional searches were not identified.

This is not ideal, but also not completely problematic. By leaving the concept to be interpreted by the LLM, at least well-known organizations were identified. Hence, there is a certain interest in leaving some concepts subject to the LLM’s interpretation, especially given the high accuracy rate for all variables. Still, further research on the topic is needed.

\subsection{Links between Variables with Similar Concepts}
During the analyses, the coders noticed some disparities between interrelated variables such as \texttt{industries}, \texttt{is\_targeting\_critical\_infrastructure}, and \texttt{is\_targeting\_large\_organization}. For example, consider this summary: 

\begin{quote}
    ``A user announced the sale of an SQL Injection vulnerability in the Peru Military Document Management System. This vulnerability allows access to documents, users, passwords, and more, with the server holding over 50 databases under the domain 'mil.pe'." (259)
\end{quote}

In this example, the LLM coded \texttt{FALSE} for \texttt{is\_targeting\_critical\_infrastructure} and \texttt{TRUE} for \texttt{is\_targeting\_large\_organization}. In the \texttt{industries} variable, the LLM indicated ['Critical Infrastructure', 'Military']. This discrepancy raises questions about the definitions of these three variables, as discussed above, and how the LLM interpreted them. It also showcases that some variables may be used to code others. For example, a technology mentioned in the summary was not always associated with an organization, despite the obvious connection between the two. In this excerpt: 
\begin{quote}
    ``A user expressed interest in purchasing generated iCloud accounts, offering a rate of \$0.5 per account" (116)
\end{quote}

The LLM indicated “iCloud” in the \texttt{targeted\_technologies} variable but did not indicate “Apple” in the \texttt{targeted\_organization} variable. Linking the two variables could enhance the information coded.

\subsection{The Title as Key}
As a reminder, the LLM system codes variables based on the conversation flows and is given some context. Such context is either the title or, if part of the conversation was already coded by the LLM, the summary of the previous conversation. The coders noticed that often, the title provided key information that, if given to the LLM, would have avoided some coding errors. For example, here is an LLM summary: 

\begin{quote}
    ``An actor expressed interest in purchasing data, with a priority on information from Israel and less emphasis on data from Iran.” (312)

\end{quote}

From this conversation, the LLM coded \texttt{is\_targeting\_critical\_infrastructure} as \texttt{FALSE}, which is accurate given the summary. However, the thread’s title is ``Buy GOV access." Hence, with this title, the LLM should have coded \texttt{is\_targeting\_critical\_infrastructure} as \texttt{TRUE} because the access the user wishes to purchase belongs to a government. Since the title was not included in the LLM's results, it missed this important detail. Potentially, adding the title with the previous summary in the context of the prompt could reduce coding errors. This relates, as well, to how the data is chunked and subsequently interpreted, as discussed above.

\subsection{The LLM System is Imperfect, just like Humans}
One last element that needs to be considered is that there were instances where the coders simply did not understand the logic behind the LLM’s coding. For example, given this summary: 
\begin{quote}
   ``A user advertised a product called 'Red Node HVNC', which is a hidden VNC (Virtual Network Computing) tool that allows remote control over a target machine via a web browser. [The actor] claims the tool is fully undetected (FUD) by antiviruses, including Windows Defender, and does not require a crypter for distribution. A demonstration video and a feature comparison chart were provided. Another user criticized the product, suggesting it is overpriced and questioning the lack of moderation for such products. [The principal actor] defended their product, stating it is privately coded, FUD, fast, and the first HVNC to work over a web browser, unlike others cloned from 'tinynuke'." (160)

\end{quote}

The LLM coded the variable \texttt{is\_initial\_access} as \texttt{FALSE} even though the actor sells a tool that allows remote control over a target machine via a web browser. Hence, in some cases, the LLM missed the information. The LLM system was therefore not flawless. Although rare, it made coding errors that could not be explained.

\section{Conclusion}
Cyber threat intelligence has traditionally been plagued by a lack of actionability and challenges around making effective use of analyst time. Our research indicates that there is significant potential for leveraging LLMs to conduct an initial review of original source data to categorize events and prioritize those that may be most relevant to CTI teams. This presents a substantial opportunity to use analyst time more effectively by narrowing searches and automatically alerting on relevant findings.

When language models go wrong, they tend to do so in predictable ways. Providing the LLM with the context needed to categorize the event was important, and it did sometimes miscategorize events, particularly when the prompt left room for interpretation. Another potential error was introduced when conversations spanned multiple days, so the summarization may have missed crucial context, thus resulting in an incorrect label. Our analysis also showed that the model struggled with differentiating between stories and original events occurring.

Despite these issues, OpenAI’s GPT-3.5-turbo model~\cite{openai_gpt3_5_turbo} was able to produce exceptional results across our study. This indicates language models are currently a scalable and cost-effective solution for identifying relevant CTI data and could have exceptional potential when paired with state-of-the-art (SOTA) models for reporting data and fine-tuning results. Leveraging SOTA models, such as Claude 3.5~\cite{anthropic_claude3_5_sonnet} Sonnet or GPT-4o~\cite{openai_hello_gpt4o, openai_gpt4_turbo, openai_gpt-4_2024}, has the potential to further improve results or allow for categorization of even more complex data. This represents a significant opportunity for further study.

\section*{Acknowledgments}
This study was supported by a mini-grant from the Human-Centric Cybersecurity Partnership (HC2P). This is the second version of the manuscript. From the first version to the second one, additional data was coded to overcome heavily unbalanced variables. The manuscript was then updated with this new data, along with recall and precision metrics. 

\bibliographystyle{plain}
\bibliography{bib}

\begin{thebibliography}{10}

\bibitem{adewopo_exploring_2020}
Victor Adewopo, Bilal Gonen, and Festus Adewopo.
\newblock Exploring {Open} {Source} {Information} for {Cyber} {Threat} {Intelligence}.
\newblock In {\em 2020 {IEEE} {International} {Conference} on {Big} {Data} ({Big} {Data})}, pages 2232--2241, December 2020.

\bibitem{anthropic_claude3_5_sonnet}
{Anthropic}.
\newblock Claude 3.5 sonnet, 2024.
\newblock Accessed: 2024-08-05.

\bibitem{ferrag_generative_2024}
Mohamed~Amine Ferrag, Fatima Alwahedi, Ammar Battah, Bilel Cherif, Abdechakour Mechri, and Norbert Tihanyi.
\newblock Generative {AI} and {Large} {Language} {Models} for {Cyber} {Security}: {All} {Insights} {You} {Need}, May 2024.
\newblock arXiv:2405.12750 [cs].

\bibitem{hassanin_moustafa_2024}
M.~Hassanin and N.~Moustafa.
\newblock A comprehensive overview of large language models (llms) for cyber defences: Opportunities and directions.
\newblock {\em arXiv}, 2024.
\newblock arXiv:2405.14487.

\bibitem{hastie2009elements}
Trevor Hastie, Robert Tibshirani, Jerome~H Friedman, and Jerome~H Friedman.
\newblock {\em The elements of statistical learning: data mining, inference, and prediction}, volume~2.
\newblock Springer, 2009.

\bibitem{liggett_dark_2020}
Roberta Liggett, Jin~R. Lee, Ariel~L. Roddy, and Mikaela~A. Wallin.
\newblock The {Dark} {Web} as a {Platform} for {Crime}: {An} {Exploration} of {Illicit} {Drug}, {Firearm}, {CSAM}, and {Cybercrime} {Markets}.
\newblock In Thomas~J. Holt and Adam~M. Bossler, editors, {\em The {Palgrave} {Handbook} of {International} {Cybercrime} and {Cyberdeviance}}, pages 91--116. Springer International Publishing, Cham, 2020.

\bibitem{openai_gpt3_5_turbo_0613}
{OpenAI}.
\newblock Gpt-3.5 turbo (0613): Function calling, 16k context window, and lower prices, 2024.
\newblock Accessed: 2024-08-05.

\bibitem{openai_gpt3_5_turbo}
{OpenAI}.
\newblock {\em GPT-3.5 Turbo Documentation}, 2024.
\newblock Accessed: 2024-08-05.

\bibitem{openai_gpt-4_2024}
OpenAI.
\newblock {GPT}-4 {Technical} {Report}, March 2024.
\newblock arXiv:2303.08774 [cs].

\bibitem{openai_gpt4_turbo}
{OpenAI}.
\newblock {\em GPT-4 Turbo and GPT-4 Documentation}, 2024.
\newblock Accessed: 2024-08-05.

\bibitem{openai_hello_gpt4o}
{OpenAI}.
\newblock {\em Hello GPT-4o}, 2024.
\newblock Accessed: 2024-08-05.

\bibitem{sheikh_artificial_2023}
Haroon Sheikh, Corien Prins, and Erik Schrijvers.
\newblock Artificial {Intelligence}: {Definition} and {Background}.
\newblock In Haroon Sheikh, Corien Prins, and Erik Schrijvers, editors, {\em Mission {AI}: {The} {New} {System} {Technology}}, pages 15--41. Springer International Publishing, Cham, 2023.

\bibitem{xu2024large}
H.~Xu, S.~Wang, N.~Li, K.~Wang, Y.~Zhao, K.~Chen, T.~Yu, Y.~Liu, and H.~Wang.
\newblock Large language models for cyber security: A systematic literature review.
\newblock {\em arXiv}, 2024.

\end{thebibliography}

\end{document}